\documentclass[twocolumn,epsf,prb,amsmath,amssymb,showpacs]{revtex4}
\usepackage{graphicx}

\begin{document}
\title{Electrostatically confined Quantum Rings in bilayer Graphene}

\author{M. Zarenia$^1$, J. M. Pereira Jr.$^2$, F. M. Peeters$^{1,2}$, and G. A. Farias$^2$}
\address{$^1$Department of Physics, University of Antwerp,
Groenenborgerlaan 171, B-2020 Antwerpen, Belgium\\
$^2$Departamento de F\'{\i}sica, Universidade Federal do Cear\'a,
Fortaleza, Cear\'a, 60455-760, Brazil.}

\begin{abstract}
We propose a new system where electron and hole states are
electrostatically confined into a quantum ring in bilayer graphene.
These structures can be created by tuning the gap of the graphene
bilayer using nanostructured gates or by position-dependent
doping. The energy levels have a magnetic field ($B_{0}$) dependence that is strikingly
distinct from that of usual semiconductor quantum rings.
In particular, the eigenvalues are not invariant under a
$B_{0} \rightarrow -B_{0}$ transformation and, for a fixed total
angular momentum index $m$, their field dependence is not parabolic,
but displays two minima separated by a saddle point. The spectra also
display several anti-crossings, which arise due to the overlap of
gate-confined and magnetically-confined states.

\end{abstract}

\pacs{71.10.Pm, 73.21.-b, 81.05.Uw} \maketitle

The electronic properties of low-dimensional systems have long
been an area of intense research. In recent years, a particularly
interesting two-dimensional (2D) system, namely single layers of crystal carbon
(graphene), has been obtained experimentally. The striking
mechanical and electronic properties of graphene have quickly been
recognized and promise to lead to new applications in electronic
devices and sensors.  These devices will benefit from the large
charge carrier mobilities and long mean free paths at room
temperature \cite{zheng,novo3,novo4,shara,zhang}.
It has been recognized that two coupled layers of graphene
(bilayer graphene - BG) is a very different material from graphene
and also from graphite. The carrier spectrum of
electrons in ideal BG is gapless and approximately parabolic at
low energies around two points in the Brillouin zone (${\mathbf K}$ and ${\mathbf K'}$)
\cite{Bart,Ohta}. In the presence of a perpendicular electric
field, however, the spectrum is found to display a gap, which can
be tuned by varying the bias, or by chemical doping of the
surface\cite{McCann}. This tunable gap can then be exploited for
the development of BG devices. In particular, the possibility of
tailoring the energy gap in BG has raised the prospect of the creation
of electrostatically defined quantum dots in bilayer graphene by
means of a position-dependent doping or through the deposition
of split gates on the BG surface \cite{Milton1}.

A very important class of quantum devices consists of
quantum rings. They have generated a lot of interest, especially because
they allow the observation of quantum
phase coherence effects on carrier transport such as in the
Aharonov-Bohm \cite{AB} and Aharonov-Casher \cite{AC} effects.
Semiconductor-based quantum rings have been obtained
experimentally \cite{Bichier} and, recently, quantum rings have
also been studied on single-layer graphene, both theoretically and
experimentally \cite{Russo,Molitor}. The latter have been produced
by lithographic techniques in which graphene nanoribbons or ring
structures are carved from an otherwise defect-free surface. Such
techniques permit the production of devices in the nanometer
scale, but have the disadvantage of creating defects at the edges
of the graphene-based structure which may reduce the overall
performance, as well as raise difficulties for the theoretical
analysis of the device. These quantum rings can also be described
as graphene flakes with a central antidot, and recent calculations
indicate that localized edge states strongly affect the
spectrum of these systems \cite{Bahamon}. Additionally, the specific
type of edge (zig-zag versus armchair) was also found to influence
their electronic properties.


To overcome the problems related to the edges of the quantum ring (i.e.
defects and type of edge) we propose here an electrostatically defined
graphene-based quantum ring (GQR). 
In contrast with previous quantum rings on graphene, the system
considered here is created by the use of electrostatic potentials
which induce a position-dependent gap, such that the low-energy
electron and hole states can be confined in a ring-shaped region
of an otherwise ideal BG sheet. This type of confinement is not
possible in single layer graphene, due to the Klein
tunneling effect \cite{Katsnelson}. In our proposed structure, the
effects of the edges are not relevant since the BG sheet is
assumed to be defect-free and the confinement is brought about by
an external bias. Moreover, in contrast with structures carved by
means of lithography, the ring parameters can be tuned by external
fields, a feature that can be particularly relevant for the design
of field configurable devices. Here, we obtain the energy levels
and wavefunctions of the confined electron and hole states by
numerically solving the Dirac equation and results are presented
as function of ring radius, width and magnetic field.

Bilayer graphene consists of two weakly, van der Waals coupled
honeycomb sheets of covalent-bond carbon atoms in a Bernal AB
stacking. The system can be described in terms of four sublattices,
labelled A, B (upper layer) and C, D (lower layer). The A and C
sites are coupled via a nearest-neighbor interlayer
hopping term $t$. We employ the continuum model based on the Dirac
equation, which is known to provide a realistic description of graphene-based
structures with dimensions that are much larger than the lattice
parameter and that has been shown to give good
agreement with experimental data. 

The Hamiltonian, in the vicinity of the ${\mathbf K}$ valley and
in the presence of a magnetic field normal to the plane of the layer
is given by \cite{McCann}
\begin{equation}
\mathcal {H}=
\begin{pmatrix}
  U_0 +\frac{\Delta U}{2}& \pi & t & 0 \\
 \pi^\dagger & U_0 +\frac{\Delta U}{2}& 0 & 0\\
  t & 0 & U_0 -\frac{\Delta U}{2}& \pi^\dagger\\
 0 & 0 & \pi & U_0-\frac{\Delta U}{2}
\end{pmatrix}
,
\end{equation}
where $t \approx 400$ meV is the interlayer coupling term, $\pi =
v_F[(p_x+eA_x) + i(p_y+eA_y)]$, ${\mathbf p}$ is the momentum
operator, $\mathbf A$ is the vector potential, $\Delta U =
U_1-U_2$ is the difference of potential between the layers,
$U_0=(U_1+U_2)/2$ is the average potential, and $v_F \approx
1\times 10^6$ m/s is the Fermi velocity. In this work we neglect
the small Zeeman splitting of the energy levels.

\begin{figure} \vspace*{0.9cm}\centering {\resizebox*{!}{6.2cm}
{\includegraphics{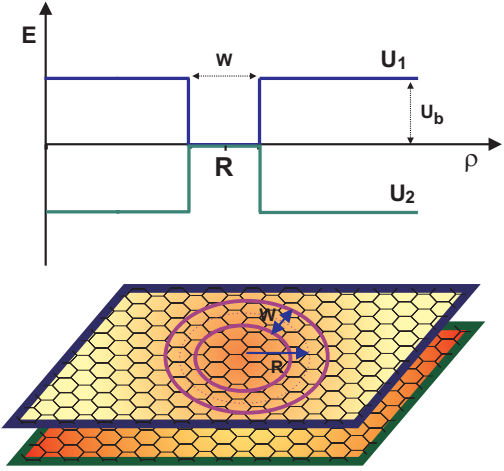}}} \caption{
Schematic depiction of the potential profile for a bilayer graphene quantum
ring}\label{fig1}
\end{figure}

The eigenstates of Eq. (1) are four-component spinors $\Psi =
[\psi_A \, , \, \psi_B\, , \, \psi_{C}\, , \, \psi_{D}]^T$, where
$\psi_{i}$ ($j=A,B,C,D$) are the envelope functions associated with the
probability amplitudes at the respective sublattice sites of the
upper and lower graphene sheets. For a circular-symmetric,
position-dependent potential such as the one described in Ref. [\onlinecite{Milton1}],
the spinor components are $\psi_A
=\phi_A(\rho)e^{im\theta}$, $\psi_B
=\phi_B(\rho)e^{i(m-1)\theta}$, $\psi_C
=\phi_C(\rho)e^{im\theta}$, and $\psi_D
=\phi_D(\rho)e^{i(m+1)\theta}$, where $m$ is the angular momentum
label. Using the symmetric gauge ${\bf A}=(0,B_{0}\rho/2,0)$, the
radial dependence of the spinor components is described, in
dimensionless units, by
\begin{eqnarray}
&&\Bigl[\frac{d}{d r'} + \frac{m}{r'} + \beta
r'\Bigr]\phi_A = -(\alpha-\delta)\phi_B,\cr &&\cr
&&\Bigl[\frac{d}{d r'} - \frac{(m-1)}{r'}- \beta
r'\Bigr]\phi_B = (\alpha-\delta)\phi_A + t'\phi_{C},\cr &&\cr &&
\Bigl[\frac{d}{d r'} + \frac{(m+1)}{r'}+ \beta
r'\Bigr]\phi_{D} = (\alpha+\delta)\phi_{C}+t'\phi_A,\cr &&\cr &&
\Bigl[\frac{d}{d r'} - \frac{m}{r'}- \beta r'
\Bigr]\phi_{C} = -(\alpha+\delta)\phi_{D},
\end{eqnarray}
where $r'=\rho/R$, $\alpha = \epsilon-u_0$, $u_0=(u_1+u_2)/2$, and
$\delta=\Delta u/2$, with $\Delta u = u_1-u_2$. The energy, the
potentials and the interlayer coupling strength are written in
dimensionless units as $\epsilon=E/E_{0}$,
$u_{1,2}=U_{1,2}/E_{0}$, $t'=t/E_{0}$ and $E_{0}=\hbar v_{F}/R$,
where $\rho$ is the radial variable and $R$ is the ring radius.
The dimensionless parameter $\beta= (eB/2\hbar)R^{2}$, can be
expressed as $\Phi/\Phi_{0}$, where $\Phi$ is the magnetic flux
threading the ring and $\Phi_{0}=h/e$ is the quantum of magnetic
flux.

In this paper we solve Eq. (2) by the finite elements method for the
following profile (see Fig. 1) :
\begin{equation}
\label{hj}
    \delta=\left\{\begin{array}{c}
                 0\qquad\qquad\qquad\qquad\qquad~ 1-\frac{w'}{2}\leq r'<1+\frac{w'}{2}\\
                 u_{b}\qquad\qquad 0<r'<1-\frac{w'}{2}~~,~~1+\frac{w'}{2}\leq r'<\infty
               \end{array}
\right.
\end{equation}
where $w'=W/R$, with $W$ being the width of the ring. This
potential describes a bilayer graphene quantum ring (GQR) of radius $R$ and width $W$, in which
both electron and holes are confined by a tunable potential
barrier $U_b$. As shown in Ref. [9], the
wavefunctions given by Eq. (2) are eigenstates of the operator
\begin{equation}
J_z = L_z + \frac{\hbar}{2}
\begin{pmatrix}
  -\mathbf{I} & 0  \\
 0 & \mathbf{I}
\end{pmatrix} +\frac{\hbar}{2}
\begin{pmatrix}
  \sigma_z & 0  \\
  0 & -\sigma_z
\end{pmatrix}
\end{equation}
with eigenvalue $m$, where $\mathbf{I}$
is the $2\times 2$ identity matrix, $L_z$ is the angular momentum operator and $\sigma_z$ is one of the Pauli matrices.
The above equation can be rewritten as $J_z = L_z + \hbar \tau_z + \hbar S_z$, where the
second term in the right hand side is a layer index operator, which is associated
with the behavior of the system under inversion, whereas $S_z$ carries the information
on the pseudospin components in each layer. It should be
emphasized that in this system, $L_z$ does not commute with the Hamiltonian and
is no longer quantized.

Figure 2 shows the lowest energy levels of a GQR
as function of ring radius $R$ (upper
panel), and width W (lower panels), for $m=-1$ (a,d), $m=0$ (b,e),
and $m=1$ (c,f), for $W=20$ nm (a-c), $R= 50$ nm (d-f),
$U_{b}=150$ meV and $B=0$. The results show a weak
dependence on the ring radius and an interesting electron-hole asymmetry,
which is due to the breaking of the inversion symmetry by the external
bias. However, the electron and hole energy levels are invariant
under the transformation $E(m) \rightarrow -E(-m)$. 
\begin{figure}
\centering{\resizebox*{!}{6.3cm} {\includegraphics{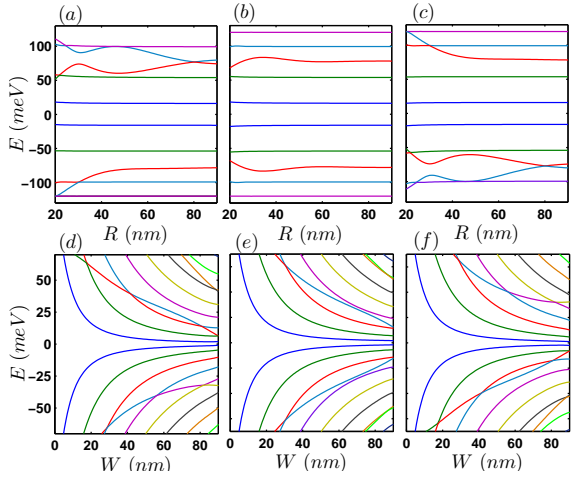}}}
\caption{
Lowest energy levels of a graphene bilayer
quantum ring as function of ring radius $R$ (Panels: $a, b,
c$) and width of the ring $W$ ($d, e, f$). Parameters given in the
text.} \label{fig2}
\end{figure}
\begin{figure}
\vspace*{0.5cm} \centering {\resizebox*{!}{7cm} {\includegraphics
{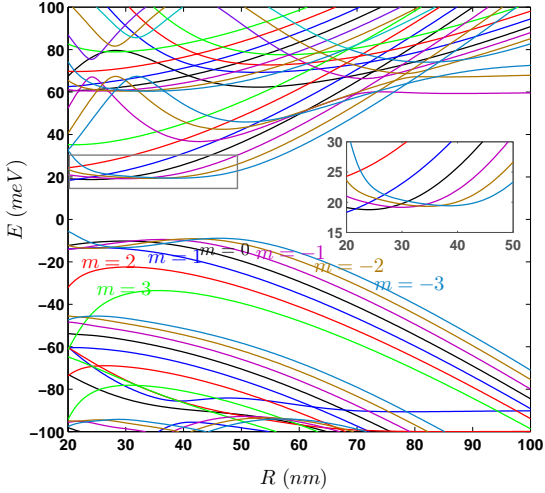}}} \caption{
Energy levels of a bilayer graphene
quantum ring as function of ring radius $R$.
Parameters given in the text.} \label{fig3}
\end{figure}
%
In Fig. 3, the lowest energy levels of a QR are shown as function
of ring radius in the presence of a non-zero magnetic field, $B_{0}=5$ T, for
$-3\leq m\leq 3$, $W=20$ nm and $U_{b}=150$ meV.
In contrast with the zero field case, the results
show a strong dependence on the ring radius, with the appearance
of several crossings, as shown more clearly in the inset. 
\begin{figure}
\vspace*{0.1cm} \centering{\resizebox*{!}{9.5 cm}
{\includegraphics{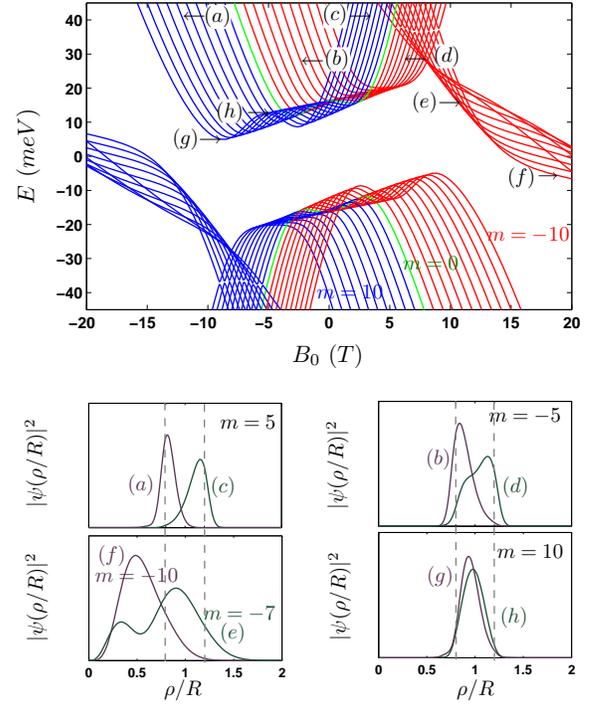}}} \caption{
Electron and hole energy states of a graphene bilayer ring-shaped quantum dot
as function of external magnetic field $B_{0}$. Parameters given
in the text.} \label{fig5}
\end{figure}

\begin{figure}
\vspace*{0.1cm} \centering{\resizebox*{!}{9.5cm}
{\includegraphics{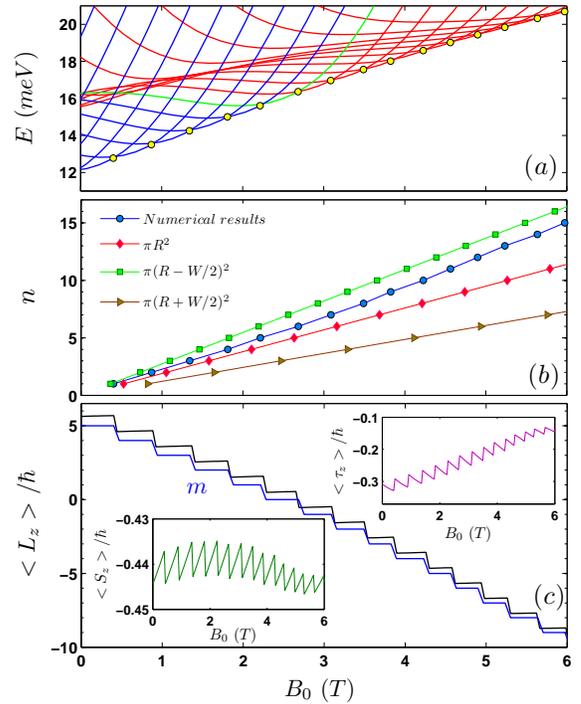}}} \caption{
$(a)$ Enlargement of the electron spectrum of Fig (\ref{fig5}). $(b)$ Number of transitions between
the levels (circles) and AB oscillations for a ring with radius $R$ and surface $S=\pi R^{2}$, $S=\pi (R\pm W/2)^{2}$. $(c)$ Ground state expectation values of $L_{z}/\hbar$, $\tau_{z}/\hbar$ and $S_{z}/\hbar$ as function of the magnetic field.}\label{fig6}
\end{figure}
The energy levels of electrons and holes for the ${\mathbf K}$
valley in a GQR are presented in Fig. 4 as function of external
magnetic field $B_{0}$, for $-10\leq m\leq-1$ (red curves), $m=0$
(green curves) and $1\leq m\leq 10$ (blue curves) with $R=50$ nm,
$W=20$ nm and $U_{b}=150$ meV. In contrast with the results
obtained from the Schr\"odinger equation, the spectra are not
invariant under the transformation $B_{0}\rightarrow -B_{0}$.
Moreover, the individual branches display two local minima. The figures
also show a set of branches with energies that decay as the field increases.
These states correspond to Landau levels of the biased bilayer,
for which the carrier confinement is mainly due to the magnetic
field. By making a small gap in the confinement region of the ring it
can be evidenced that those branches are only weakly affected by a variation
of the gap. We found that the electron and hole
energies are related by $E_e(m,B_0)= -E_h(-m,-B_0)$. The spectrum
for the ${\mathbf K'}$ valley can be obtained by setting
$B_{0}\rightarrow -B_{0}$, which shows that, as in the case of the
graphene bilayer quantum dot \cite{Milton1}, the valley degeneracy
is lifted. The four lower panels show the probability density of
the points on the spectrum which are labeled by $(a),(b),...$ in
the upper panel. The vertical dashed lines indicate the edges of the ring.
As seen, for the (e) and (f) points the probability density has
peaks in the internal region of the ring structure ($\rho < R - W/2$), whereas for the remaining
points the carriers are mainly confined inside the ring. The (e) and (f) states belong to energy states
that in the limit $B_{0}\rightarrow\infty$ approach the zero Landau level are shifted from $E=0$ due to the presence of the ring potential.
An enlargement of the region of the spectrum is
shown in Fig. 5(a). The yellow dots correspond to the location of the
transitions of the ground state between the levels which is related to the number of
AB oscillations $n=\Phi/\Phi_{0}$.
Figure 5(b) shows the number of
transitions between the levels (circles) and the AB oscillations for a
zero-width ring with radius $R$ and surface $S=\pi R^{2}, S=\pi (R\pm W/2)^{2}$.
Notice that the numerical results obtained for the finite width ring follow closely
the AB oscillations as expected for an ideal ring with radius equal to the one of the
inner ring.

The ground state expectation values of the $L_{z}/\hbar$, $\tau_{z}/\hbar$ and
$S_{z}/\hbar$ operators are shown as function of the magnetic field
in Fig. 5(c). The figure shows that the values of $<L_{z}>/\hbar$ (black curve), $<\tau_{z}>/\hbar$
(upper inset) and $<S_{z}>/\hbar$ (lower inset) are not quantized, but,
their sum, as in Eq. (4) (blue curve) is given by the $m$ quantum number. Very different from the
Schr\"odinger case we find that for $B_{0}\approx0$ the quantum number $m$ and the average $<L_{z}>$ are
non-zero.

The spectrum in
Fig. 4 exhibits several anticrossings between different branches. An
enlargement of one particular anticrossing (between the first and second level of $m = -1$)
is displayed in Fig. 6(a). The corresponding values
of $<L_z>$, the average radial position $<\rho/R>$, as well as the
probability density for the points (1), (2) and (3) are shown
in panels (b), (c) and (d), respectively. The figure indicates
that at the anti-crossing the eigenstate evolves from a
configuration in which the electron is mainly confined in the
center of the ring by the magnetic field, to a state in which the
confinement is mainly caused by the electrostatic potential $U_b$.
At the anti-crossing (2), the probability density plot shows
peaks both in the central part of the structure, as well as in the
ring, which characterizes the overlap of the magnetically and
electrostatically confined states.
\begin{figure}
\vspace*{0.5cm}\centering{\resizebox*{!}{9.2cm}
{\includegraphics{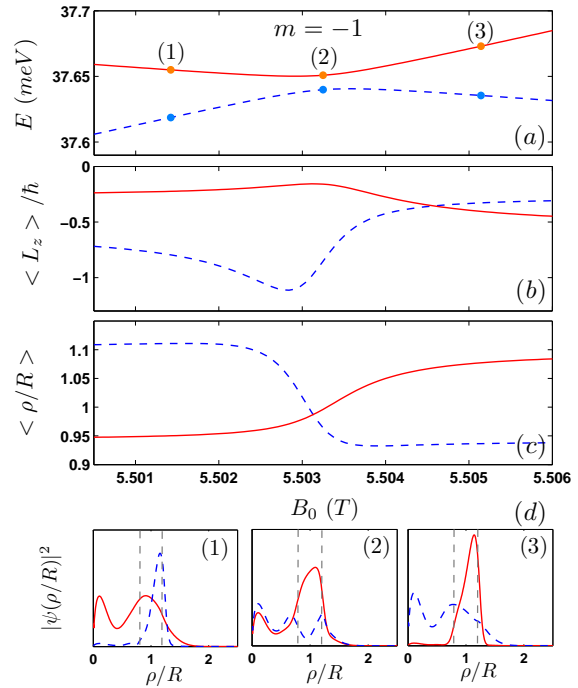}}} \caption{$(a)$ Enlargement around
an anti-crossing point in the energy spectrum. $(b)$ Expectation value of angular momentum
$<L_{z}>/\hbar$ and $(c)$ radial average position $<\rho/R>$
around the anti-crossing point shown in $(a)$. $(d)$ Probability
density for the points $(1)$, $(2)$ and $(3)$ shown in $(a)$.}
\label{fig7}
\end{figure}
%

In summary, we obtained the eigenstates and energy levels of
electrostatically confined quantum rings in graphene bilayers.
This new approach differs from previous studies of graphene-based
quantum rings which were realized through a lithographic cutting of
the sample. The advantage of the present approach is that the confinement
can be tuned by varying the profile and depth of the confining
potential, preventing issues as edge disorder and the specific edge type.
Numerical results were presented for a step-like
confining potential, which can be expected to provide a good
approximation of actual potential barriers, because of the large
screening in bilayer graphene. We predict that the energy
levels display a dependence on the magnetic field that is quite
distinct from that of a conventional 2D electron gas. Moreover,
the energy levels are not invariant under a $B_{0}\rightarrow
-B_{0}$ transformation, in contrast with usual ring
structures. This is a consequence of the fact that the ring
structure is produced by a gate that introduces an electric field
and thus a preferential direction. On the other hand, the spectrum
is found to be invariant under the transformation
$B_{0}\rightarrow -B_{0}$, together with $U_{b}\rightarrow -U_{b}$ and
$m\rightarrow -m$. Alternatively, $B_{0}\rightarrow -B_{0}$
and $E\rightarrow -E$ transforms electron states into
hole states and vice-versa. The spectra also present several
anti-crossings at low energies, which arise due to the overlap of
gate-confined and magnetically-confined states. For a fixed total
angular momentum index $m$, the $E(B_{0})$ curves are no longer
parabola, but show two minima separated by a saddle point. The
existence of Aharonov-Bohm oscillations for both electrons and
holes are still linked with flux quantization through the ring.
Throughout the calculation, we considered single electron states.
However, the inclusion of Coulomb interaction between carriers may
introduce interesting modifications in the spectrum \cite{Abergel}. The system
can be realized experimentally by a suitable choice of nanostructured doping
levels or with the application of nanostructured gates. The spectra
may be investigated by, e.g. cyclotron resonance methods
\cite{Kim} and quantum transport measurements.

{\bf{Acknowledgements}}.
 This work was supported by the Flemish Science Foundation
(FWO-Vl), the Belgian Science Policy (IAP), the Bilateral program between Flanders and Brazil,
and the Brazilian Council for Research (CNPq).

\end{document}